\documentclass[12pt,preprint]{aastex}

\slugcomment{Submitted to: } \shorttitle{Orbital parameters and
pulse evolution of V0332+53 in outburst} \shortauthors{ZHANG et
al.}

\begin{document}
\title{Recovery of the orbital parameters and pulse evolution of V0332+53 during a huge outburst}

\author{Shu Zhang$^1$, JinLu Qu$^1$, LiMing Song$^1$ and Diego F. Torres$^2$ }
\affil{$^1$Laboratory for Particle Astrophysics, Institute of High
Energy Physics, Beijing 100049, China,E-mails:szhang@mail.ihep.ac.cn, qujl@mail.ihep.ac.cn,songlm@mail.ihep.ac.cn
%\affil{et}
%\email{szhang@mail.ihep.ac.cn}
\and
$^2$Lawrence Livermore National Laboratory, 7000 East Ave., L-413,
Livermore, CA 94550, E-mail: dtorres@igpp.ucllnl.org}

\begin{abstract}
The high mass X-ray binary (HMXB) V0332+53 became active at the
end of 2004 and the outburst was observed at hard X-rays by RXTE and
INTEGRAL. Based on these hard X-ray observations,
the orbital parameters are
measured through fitting  the Doppler-shifted spin periods.
The derived orbital period and 
eccentricity are consistent with those of Stella et al.
(1985) obtained from  EXOSAT observations, whereas the projected
semimajor axis and the periastron longitude are found to have
changed  from 48$\pm$4 to 86$^{+6}_{-10}$ lt-s and from
313$^{\circ}$$\pm$10 to 283$^{\circ}$$\pm$14, respectively.
This would indicate an angular speed  of $\geq$ 1.5$^{\circ}$$\pm$0.8
yr$^{-1}$ for rotation of the orbit over the past 21 years. The periastron
passage time of MJD 53367$\pm$1 is just around the time when the
intensity reached  maximum and an orbital period earlier is
the time when the outburst started. This correlation resembles
the behavior of a  Type I outburst.
During outburst the source spun up with a rate of
8.01$^{+1.00}_{-1.14}$$\times$10$^{-6}$ s
day$^{-1}$. The evolution of pulse profile is highly intensity
dependent. The separation of double pulses remained almost
constant ($\sim$ 0.47) when the source was bright, and dropped to
0.37 within $\leq$ 3 days as the source became weaker. The pulse
evolution of V0332+53 may correlate to the change in dominance
of the emission between fan-beam and pencil-beam mechanisms.

\end{abstract}


\section{INTRODUCTION}

Four outbursts of HMXB V0332+53 were recorded at hard X-rays. The
first one was a Type II outburst caught by Vela 5B in 1973, with a
duration of $\sim$ 100 days and peak intensity of 1.6 Crab
(Terrell and Priedhorsky 1984).
 A spin period of 4.37 s and an orbital period of 34 days were discovered
(Whitlock 1989).
According to the classification of 
outburst in Stella et al. (1986), this Type II outburst was an irregular
transient activity with timing unrelated to any underlying orbital period.
 The second outburst was detected by Tenma
satellite 10 years later (Tanaka et al. 1983). The follow-up
series of observations performed by EXOSAT showed the outbursts
were Type I -- periodic transient activities with
occurring time close to the time of periastron passage -- and
measured the orbit with a moderate
eccentricity of 0.31$\pm$0.03 and a projected semimajor axis of
48$\pm$4 lt-s (Stella et al. 1985). Favored by the precise
source location attained with EXOSAT, the companion was identified
in the optical as an early-type star, BQ Cam (Honeycutt and
Schlegel 1985), and distance to the source was estimated to be
2.2--5.8 kpc (Corbet et al. 1986).  V0332+53 was found to be
active again by Ginga in 1989 (Tsunemi et al. 1989). The outburst
was classified as Type II and a feature of 0.05 Hz QPO was
reported (Takeshima et al. 1994). A cyclotron absorption component
observed at 28.5 keV suggested that the magnetic filed could be as high as
2.5$\times$10$^{12}$ G on surface of the neutron star
(Makishima et al. 1990).

The most recent outburst was detected by the all sky monitor (ASM)
of RXTE in November 2004, followed by INTEGRAL
 Target of Opportunity (ToO) observations
 at hard X-rays. Three cyclotron
lines detected  in INTEGRAL  (Kreykenbohm et al. 2005)
 confirmed the earlier report in RXTE (Coburn et al. 2005).
 The intensity averaged on  December 22 were about 1.2 Crab in
the 1.5--12 keV band and 2.2 Crab in the 5--12 keV band. An
additional feature of 0.22 Hz QPO  was discovered from  PCA/RXTE to
ride on the spin frequency (Qu et al. 2005).  In this letter we
report the measurement of current orbital parameters 
 with RXTE/INTEGRAL data and the trend of pulse evolution
with time.

\section{OBSERVATIONS AND ANALYSIS}

The public data we have analyzed are shown in Figure 1 both for
PCA/RXTE and INTEGRAL observations of V0332+53 during the
outburst. The PCA/RXTE data are collected within time periods of
December 28-30 in 2004, January 5-6, 15-19, February 12-15, and
March 7 in 2005. The typical exposure of individual observation
are several thousand seconds and the neighboring observations are
combined to have the largest time span of roughly half day. The
arrival time is corrected to the barycenter for each photon (2-60
keV).

INTEGRAL ToO observations are available for 9 revolutions during
January -  February 2005. ISGRI/IBIS data of revolutions 272-274
(pointings 75, 66 and 82) ($>$ 15 keV) are adopted  in our
analysis.  The source intensity went down thereafter and the
later observations do not  allow  precisely  deriving
the spin period. The revolution 272 was carried out in staring
mode and the other two revolutions in hexagonal dithering mode
(for definitions see Chernyakova  2004). The total exposure
considered are about 110 ksec for these three revolutions. To
further improve the statistics, the neighboring science windows of 
pointing in each
 revolution are combined
to enlarge the time span to $\sim$ 10$^4$ seconds.
Data reduction is performed using the latest
version (4.2) of the standard INTEGRAL Offline Science Analysis
(OSA) software. The photons with pixel illuminated factor 1 are
extracted and the arrival time is corrected to the barycenter
(Chernyakova  2004).

\section{RESULTS}

\subsection{Orbital parameters}

The orbital parameters of binary system can be measured
through fitting the radial-velocity curve. The radial velocity
($v_r$) of orbital motion will modulate the spin period
($P_{spin}$) of  neutron star through the Doppler effect as
$P_{obs}=(P_{spin}+\dot{P}_{spin}\Delta t)\sqrt{1+v_r/c\over
1-v_r/c}$, where $\dot{P}_{spin}$ is the change rate of spin
period and $P_{obs}$ the observed spin period. For 
representation of $v_r$ by the orbital parameters see Hilditch
(2001). The systemic velocity is not considered since the $v_r$
profile is not modulated with a constant velocity. Finally we have
7 fundamental parameters to be inferred from fitting the
$P_{obs}$ curve.

The spin period is searched in observational data by using 
 $efsearch$ of the ftool software package. The resolution and
the number of phase bins ($N_b$) are chosen as 1$\times$10$^{-6}$
s and 200, respectively. The derived $\chi^2 \sim P_{obs}$ curves
are fitted by Gaussians, to estimate the most likely value of
$P_{obs}$.\footnote{The last science window of pointing 75 in
revolution 272 of ISGRI/IBIS and the PCA/RXTE data on March 9 2005
are not used since the resulting distributions in spin period are
quite noisy.} Its error is roughly taken to the first order as
$P_{obs}^2/(T_{span} \, N_b)$, where $T_{span}$ is the total time
of observation, varying from 10$^3$ to 10$^4$ s.  We note that the
variation of $v_r$
 within $T_{span}$ is
$\Delta v_r=-K_x \sin(\theta+\omega)\, \dot{\theta}\, T_{span}$,
where $\dot{\theta}$ = $2\pi\, (1-e^2)^{1/2}/(P_{orbit}(1-e \,
\cos(E))^2)$ and $E$ is the eccentric anomaly. For the first trial
in the fitting, the uncertainty in $v_r$ is taken as
2$\pi$$K_{x}T_{span}$/$P_{orbit}$, i.e. with zero eccentricity
and other parameters coming from Stella et al. (1985). The corresponding
contributions to the error in $P_{obs}$ are then taken into
account accordingly.\footnote{We have performed other trials as
well, with different combinations of Stella's results. All
obtained fits were worse than the one we find as a solution.}

The reduced $\chi^2$ for fitting the data is 0.98 (26 dofs). The
best-fit parameters for eccentricity $e$ and orbital period
$P_{orbit}$ are
consistent within  2 $\sigma$ error bar
with those of Stella et al. (1985). The projected
semimajor axis $a_{x} \sin(i)$ is obtained as 86$^{+6}_{-10}$
lt-s, which is much larger than the previous 48$\pm$4 lt-s
(Stella et al. 1985). With the updated
orbital parameters,  contributions from variation of $v_r$
 to the error in $P_{obs}$ are then revised. The
obtained spin periods from ISGRI/IBIS and PCA/RXTE are shown in
Figure 2. The data are fit again and the best-fit parameters are
shown in Table 1. The reduced $\chi^2$ remains almost unchanged
 (26 dofs).  The 1 sigma errors are obtained by adding 8.1 to the
minimum $\chi^2$-value as is appropriate for 7 parameters of
interest. The results are consistent within error bars with those
obtained 21 years ago for those parameters of $e$, $P_{orbit}$,
and $P_{spin}$, but not for $a_{x} \sin(i)$ and 
periastron longitude $\omega$. The neutron star passed
through periastron at the time when the source had  peak
intensity. We find that the neutron star spun up with a rate of
8.01$^{+1.00}_{-1.14}$$\times$10$^{-6}$ s/day during the outburst.

\subsection{Phase evolution}

To obtain the absolute phase  one has to
subtract the effect of orbital motion from the photon arrival
time. The main uncertainty in such correction comes from that in
$(a_x \, \sin(i)/c)$, which could be as large as 6-10 seconds in
our data and is insufficient for precisely obtaining the absolute
phase at different time. We therefore co-align the phase by
taking one of the two bridges for double-pulse light curves but
use arbitrary phase for light curve with single pulse (Figure 3).
As shown in Figure 1, there are five data groups representing
different intensity levels of the outburst. For each time period
the typical light curve is plotted in Figure 3. The light curves
are dominated by broad, asymmetric double pulses when the source
was at high intensity level within MJD 53367-53376. The two pulses
become narrower and symmetric as the source intensity went down to
middle level in MJD 53386. Roughly 30 days later, when the
outburst evolved to the tail, the two pulses started to move
closer. As shown in Figure 1, phase separation of the two
pulses remained roughly constant ($\sim$ 0.47) when the source was
bright, and dropped from 0.47 to 0.37 within less than 3 days (MJD
53413-53416) at a low intensity level.  In MJD 53436 the outburst
almost ceased and only a single Gaussian-shaped pulse appears in
the light curve.  The single pulse precludes estimating phase
separation in the light curve. Nevertheless, a zero phase
separation would be consistent with the trend as seen in the
preceding data (see dotted line in Figure 1).

\section{DISCUSSION}

Some orbital parameters recovered from PCA (RXTE) and ISGRI/IBIS
(INTEGRAL) observations on the recent outburst of V0332+53 differ
from those of Stella et al. (1985). To investigate such
differences in detail we tracked the procedure of parameter
measurements in Stella et al. (1985), where the orbital parameters
were estimated based on fitting  nine values of $P_{obs}$. To
fully repeat their work is difficult because only five values of
$P_{obs}$
 were presented in their paper. Fortunately, the
outburst was also monitored by  Tenma during 1983-1984, and
additional five values of $P_{obs}$ were published  in Makishima
et al. (1990). By fitting these ten data points, the orbital
parameters  are obtained with values consistent with those of
Stella et al. (1985). The spin-up rate $\dot{P}_{spin}$ is set to zero as well,
i.e. to follow
 Stella et al. (1985), and the reduced $\chi^2$ is  0.85.
Cross checking on the consistency of data with different sets
of orbital parameters is also performed. With
$\dot{P}_{spin}$ being free, both fitting  the EXOSAT/Tenma data
with the parameters  derived from RXTE/INTEGRAL and fitting
RXTE/INTEGRAL data with the parameters of Stella et al. (1985)
result the reduced $\chi^2$ far beyond 1. This provides evidence
for a change in the orbital parameters since 1984.
 Stella et al. (1985) reminded that their orbital
parameters  are only valid under very small value of $\dot{P}_{spin}$,
%(less than 3$\times$10$^{-12}$ s s $^{-1}$),
 the contemporary value of which was however unknown.
Through fitting the EXOSAT/Tenma data, we find also that
$\dot{P}_{spin}$ can not be constrained: very good fits could be
established under quite different values of $\dot{P}_{spin}$. This
may due to the relatively poor coverage of the orbital phase with
only 10 data points.  The orbital phase coverage is now improved
with as much as 33 data points originated from the RXTE/INTEGRAL
observations and the fit  can only be acceptable by including
$\dot{P}_{spin}$. Therefore, we tend to believe that the orbital
parameters  obtained from RXTE/INTEGRAL  have less uncertainty
compared to those of Stella et al. (1985).

The inconsistency of the combined EXOSAT/Tenma data  with the
current orbital parameters suggests that at least some of the
orbital elements have  changed in the past 21 years. In case that the
orbital parameters of Stella et al. (1985) are appropriate as
well, i.e.  assuming the contemporary value of $\dot{P}_{spin}$
was very small,  $a_{x} \sin(i)$ should have increased from 48$\pm$4 to
86$^{+6}_{-10}$ lt-s and $\omega$ decreased from
313$^{\circ}$$\pm$10 to 283$^{\circ}$$\pm$14. This would then imply a
rotation of the orbit with an angular velocity $\geq$
1.5$^{\circ}$$\pm$0.8 yr$^{-1}$. The ratio of $a_{x} \sin(i)$
between Stella's results and ours is 1.79$\pm$0.25. By assuming no
change in semimajor axis and inclination angle, this value is
consistent within 1 $\sigma$ error bar  with the ratio
1.34$\pm$0.22 represented by the periastron longitude. Such
consistency lies in the fact that any rotation of the orbit would
lead to subsequently changing in  the projection of semimajor
axis  along the line of sight.
 The rotation of  orbit in its own plane,  usually
defined as apsidal
motion, was widely detected in  binary systems of different configuration
(Petrova and Orlov, 1999). Apsidal motion can be induced by a variety of
effects in perturbing gravitational potential, including the rotational potential, the tidal potential
as well as the general-relativity corrections to Newtonian gravitational theory. For an
idealized Newtonian binary system, the period of apsidal motion (P$_{A}$) is
$P_{A}=P_{orbit}\left[15k_{1}\left(\frac{R_{1}}{a}\right)^{5}\frac{M_{2}}{M_{1}}\frac{1+1.5e^{2}+0.125e^{4}}{(1-e^{2})^{5}}\right]^{-1}$ (Schwarzschild 1958).
Here $k_{1}$ is the apsidal constant, $a$ the
semimajor axis of the relative orbit and the contribution from  neutron star
is neglected. With $k_{1}$ $\sim$ 0.008 for a main-sequence
 star (Stothers 1974),  a companion of $M_{1}$ $\sim$ 20
   $M_\odot$ and a neutron star of  $M_{2}$$\sim$ 1.44 $M_\odot$, the apsidal period
is estimated to be $\sim$ 8.6$\times$10$^5$ yr, corresponding to
an angular velocity $\dot{\omega}$ $\sim$
4$^{\circ}$$\times$10$^{-4}$ yr$^{-1}$. The contribution from 
general-relativity corrections  is $\dot{\omega}=\frac{6\pi
G(M_{1}+M_{2})} {P_{orbit}c^{2}a(1-e^{2})}$ (Hilditch 2001).
$\dot{\omega}$ is then estimated as $\sim$
5$^{\circ}$$\times$10$^{-3}$ yr$^{-1}$
 for V0332+53. One sees
that an angular velocity  $\sim$ 1.5$^{\circ}$$\pm$0.8 yr$^{-1}$
can not be accounted for from the contributions  of both effects.
Apsidal motion could also be an intrinsic phenomenon for a
triple system. It is recognized that roughly 20 percent of binary
system are members of triple or multiple systems. The argument of
apsidal motion in a triple system depends on the so-called
mass-distance relation (Wilson and Fox 1981). An adjustment on 
mass and  distance of the desired third body can induce large
$\dot{\omega}$, e.g. Wilson and Fox (1981) explained the rapid
apsidal period of 4.5 yr for the HMXB Cyg X-1 in terms of a
perturbing third body. The apsidal motion of $\dot{\omega}$ $\sim$
1.5$^{\circ}$$\pm$0.8 yr$^{-1}$ in V0332+53, if it is true, may
need to be understood in a scheme of triple system.

For  compact system with small orbital period (typically less than 1 day), the angular
velocity of the apsidal motion could be as large as several degrees -- hundred
degrees per year (Petrova and Orlov, 1999). For X-ray binary system  with
long orbital period, the largest rate of apsidal motion  observed so far
may be in Vel X-1 (orbital period 8.97 days) with an angular velocity of
6.9$^{\circ}\pm$3.4 yr$^{-1}$ (Boynton et al. 1986). A rate of apsidal motion
 very close to V0332+53 was reported for
SS433 (orbital period 13.08 days) with an angular speed of
2.45$^{\circ}$$\pm$0.44 yr$^{-1}$ (Petrova and Orlov, 1999).

With a mass function of
0.58 M$_\odot$, binary system V0332+53 consisting of a companion of $\geq$20
M$_\odot$ and a neutron star of 1.44 M$_\odot$ would require an
inclination angle of the orbit to be $\leq$ 18.9$^{\circ}$. With such an
inclination angle, the rotational velocity of the companion O-type star
is then estimated as 480 km s $^{-1}$ --  well below the
break-up velocity 600 km s $^{-1}$ typical for a late O-type star
(Negueruela et al. 1999). Therefore,
the evidence once reported by
Negueruela et al. (1999) for a tilt between  orbit and equatorial plane
does not hold any more under the measurements of current orbital parameters.

The periastron passage time ($T_p$) of  MJD 53367$\pm$1 is just around the
time when the intensity reached  maximum and an orbital period
earlier is the time when the outburst started ($\sim$MJD 53332,
Swank et al. 2004). This correlation resembles the behavior  of a
Type I outburst. During the outburst a spin-up rate of
8.01$^{+1.00}_{-1.14}$$\times$10$^{-6}$ s
day$^{-1}$ is derived. Such change in pulse period is generally
regarded as the result of interaction of  neutron star with
 accreting matter. The corresponding relationship between
$\dot{P}_{spin}$/$P_{spin}$ and luminosity is
$\dot{P}_{spin}$/$P_{spin}$$\sim$-3$\times$10$^{-5}$$P_{spin}$$L^{6/7}$
yr$^{-1}$ (Rappaport and Joss 1977). Here $P_{spin}$ and $L$ are
in units of second and 10$^{37}$ erg s$^{-1}$, respectively. With
the updated orbital parameters the luminosity is estimated to be
$\sim$5$\times$10$^{37}$ erg s$^{-1}$. This value is consistent
with the luminosity of (1.4-9.7)$\times$10$^{37}$ erg s$^{-1}$  in
2-10 keV during December 24-26 2004, estimated by assuming that the
source is at a distance of 2.2--5.8 kpc (Soldi et al. 2005).

It is generally thought that XRB outbursts are powered by 
accretion of the matter from the companion to the magnetic poles
of neutron star. For luminosity $\geq$10$^{37}$ erg s$^{-1}$ the
accretion flow onto the magnetic pole will be decelerated in a
radiative shock above the neutron star surface (Basko and Sunyaev
1976). The photons will escape in a fan-beam from the emission
region below the shock, and the peak emission is
perpendicular to the magnetic axis.  For luminosity lower than
10$^{37}$ erg s$^{-1}$, a pencil-beam emission might be formed
with maximum along the direction of the magnetic axis.
During the outburst, emission from both components may contribute
to the pulse profiles in the light curve. The study on  pulse
profile evolution of EXO 2030+375 indicated that, as the
luminosity decreased, the dominant emission changed from a
fan-beam to a pencil-beam configuration (Parmar et al. 1989). The
pulse evolution in V0332+53 may be understood in a similar way. At
high luminosity, the pulse profile is dominated by the fan-beam
component, showing double peaks separated in phase by $\sim$0.47.
The luminosity at MJD 53416 went down to $\sim$ 1$\times$10$^{37}$ erg
s$^{-1}$ and the contribution from an additional component, that
of the pencil-beam, showed up in light curve at somewhere between
the double pulses. At even lower intensity level, only pencil-beam
emission persists, generating a single symmetric pulse in the
light curve either because of a special geometrical configuration
of the two magnetic poles or due to cease of the flare in one of the
magnetic poles. The combination of the single pulse from pencil beam to one
of the double pulses from fan beam will lead to a broadened pulse and
the visual phase separation of the double pulses becomes smaller than previous.
That the pulse separation decreased by 20 percent,
from 0.47 to 0.37 within $\leq$ 3 days, may constrain time scale of
the transition in emission between different components when the
burst evolved to a low intensity level.

\acknowledgments This work was subsidized by the Special Funds for
Major State Basic Research Projects and by the National Natural
Science Foundation of China. The work of DFT was performed under
the auspices of the U.S. D.O.E. (NNSA), by the University of
California Lawrence Livermore National Laboratory under contract
No. W-7405-Eng-48.

\clearpage
\begin{figure}[b]
\epsscale{0.80}
\plotone{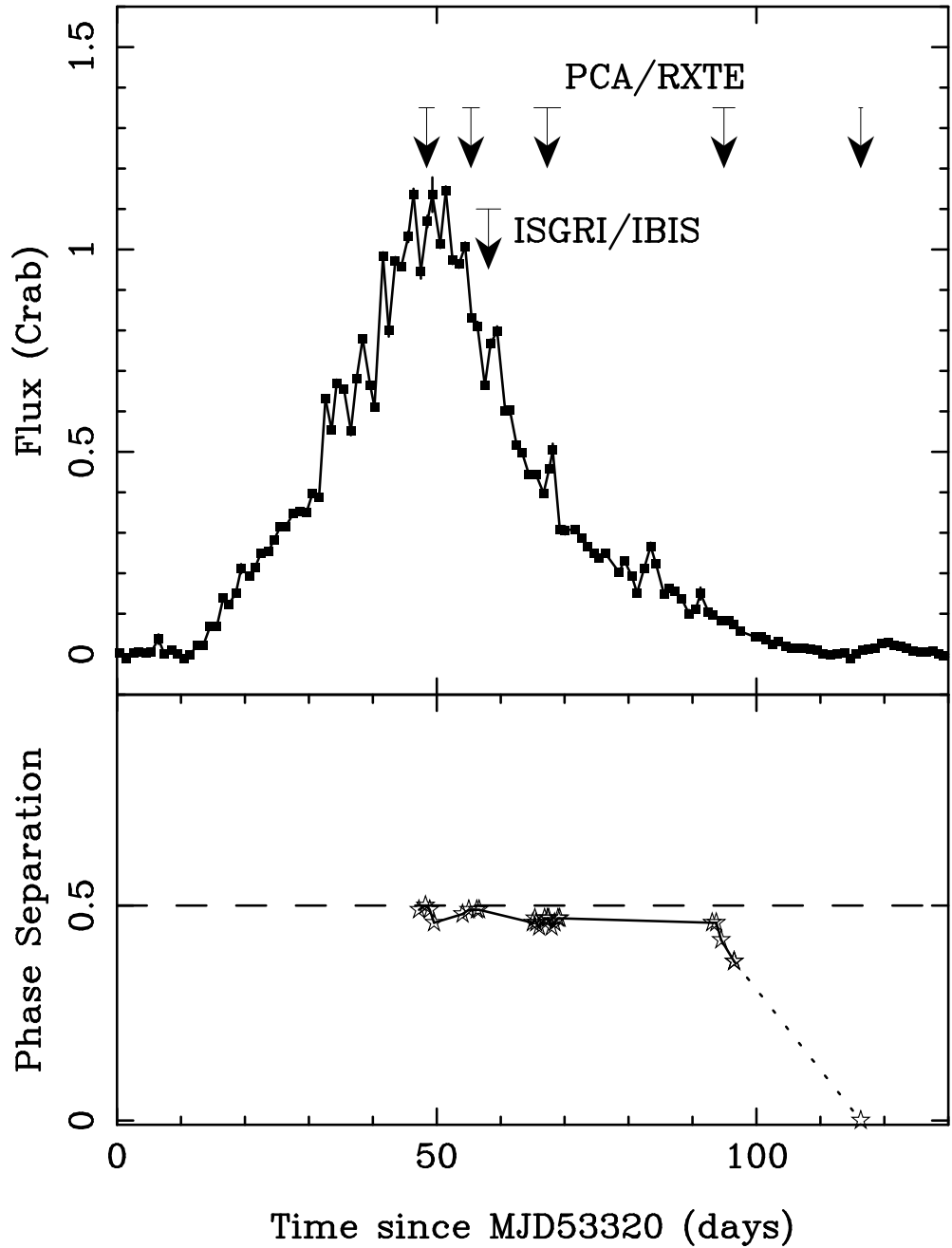}
\caption{ASM light curve (1.5-12 keV, upper panel) and  phase
separation of the pulses in light curve of V0332+53 during
outburst (lower panel). On top of the light curve in the upper
panel are over plots for PCA/RXTE and INTEGRAL observations
analyzed in this paper. The dotted line in the lower panel shows
the trend of extending to a zero phase separation.}\label{lc-phase}
\end{figure}

\clearpage
\begin{figure}[t]
\epsscale{0.80}
\plotone{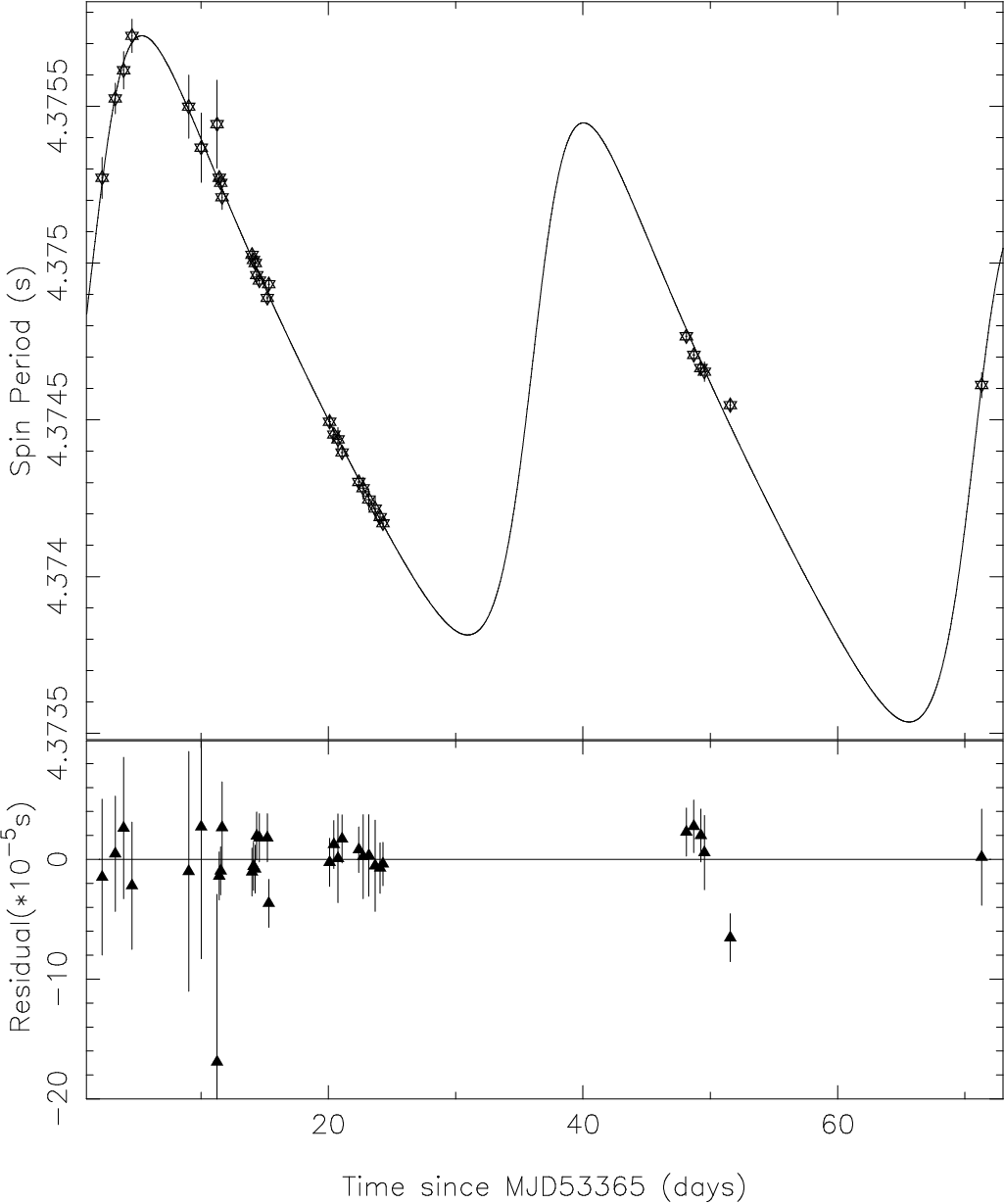}
\caption{Doppler effect on the spin period (the symbols of the
stars in the upper panel) and the curve with the best-fit orbital
parameters (upper panel). The corresponding residuals are shown in
the lower panel.}\label{fit-spin}
\end{figure}

\clearpage
\begin{figure}[t]
\epsscale{0.80}
\plotone{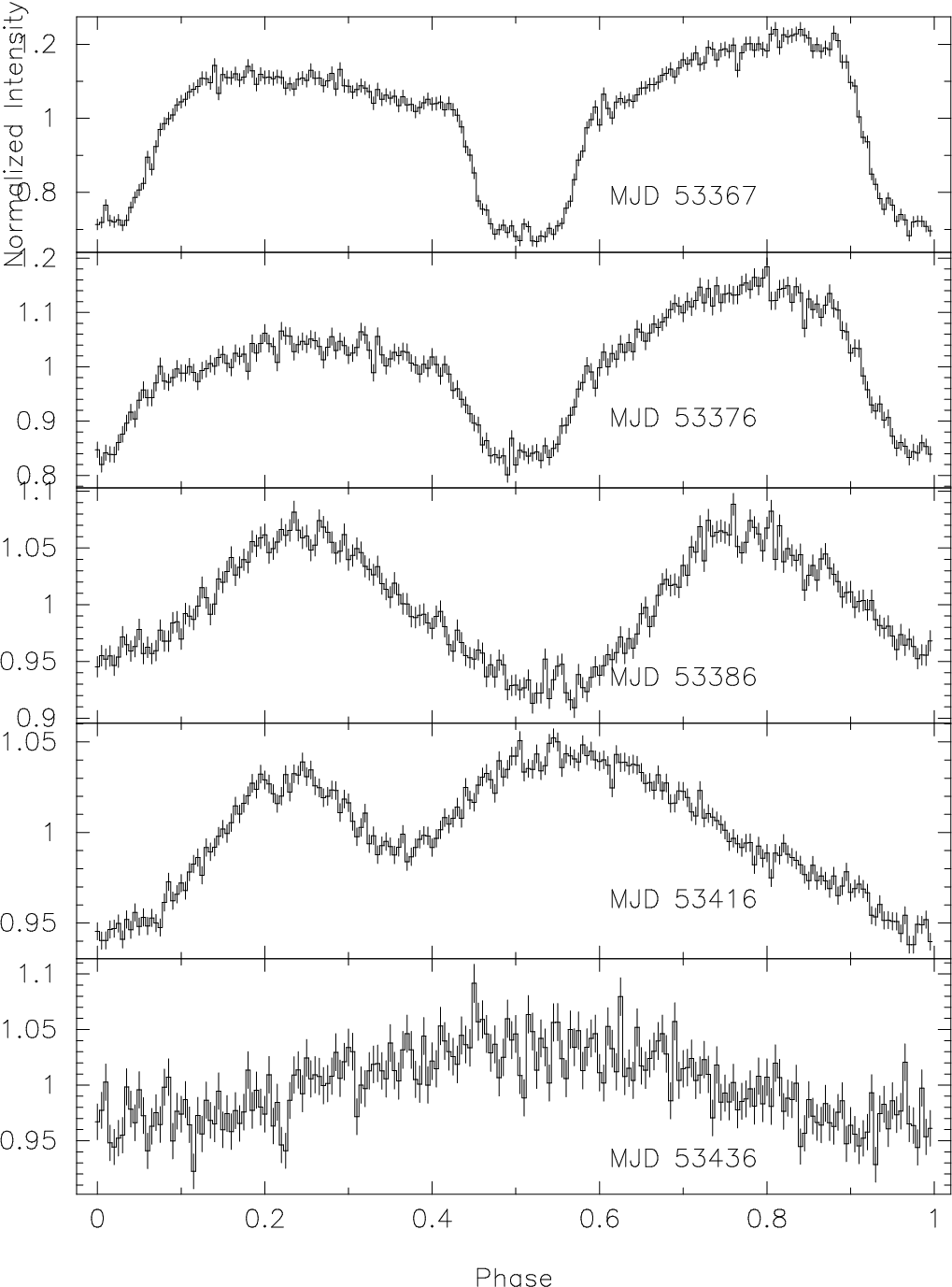}
\caption{Typical light curves at different intensity level of
V0332+53 during the outburst. }\label{lc-time}
\end{figure}

\clearpage
 %%%%% Table 1 %%%%%%%%%%%
\begin{table*}[t]
\caption[]{Parameters for V0332+53}
%\begin{flushleft}
\begin{tabular}{cc}
\hline\noalign{\smallskip} \hline\noalign{\smallskip}
$a_{x}\, \sin(i)$ & 86$^{+6}_{-10}$ lt-s  \\
$P_{spin}$ & 4.37480$^{+9\times10^{-5}}_{-5\times10^{-5}}$ s  \\
$P_{orbit}$ & 34$^d$.67$^{+0.38}_{-0.24}$   \\
$K_x$ & 58.5 $\pm$7.3 km s$^{-1}$  \\
$f(M)$ & 0.58 $\pm$ 0.23$M_{\odot}$\\
$e$ &  0.37$^{+0.11}_{-0.12}$\\
$\omega$ & 283$^{\circ}$$\pm$14\\
$T_p$ & MJD 53367 $\pm$ 1\\
$\dot{P}_{spin}$& -8.01$^{+1.00}_{-1.14}$$\times$10$^{-6}$ s/day\\
$\dot{P}_{spin}$/$P_{spin}$ & -(1.83$\pm$0.23)$\times$10$^{-6}$ yr$^{-1}$\\
\hline\noalign{\smallskip}
\end{tabular} %\end{flushleft}
\label{tab1}
\end{table*}
%%%%% Table 1 %%%%%%%%%%

\end{document}